\documentclass[12pt]{article}
\usepackage{amsmath,epsf,amssymb,latexsym,cite,pifont}

\newtheorem{theorem}{Theorem}
\newtheorem{prop}{Proposition}

\newif\iffigs\figstrue

\DeclareFontFamily{U}{rsf}{}
\DeclareFontShape{U}{rsf}{m}{n}{
  <5> <6> rsfs5 <7> <8> <9> rsfs7 <10-> rsfs10}{}
\DeclareMathAlphabet\Scr{U}{rsf}{m}{n}

\def\O{\Scr{O}}

\def\P{{\mathbb P}}
\def\Q{{\mathbb Q}}
\def\R{{\mathbb R}}
\def\Z{{\mathbb Z}}

\def\Im{\operatorname{Im}}

\def\Area{\operatorname{Area}}
\def\Vol{\operatorname{Vol}}

\def\Pic{\operatorname{Pic}}

\def\Sl{\operatorname{SL}}
\def\Gl{\operatorname{GL}}
\def\GO{\operatorname{O{}}}

\def\GU{\operatorname{U{}}}

\def\Span{\operatorname{Span}}

\def\KIII{\mathrm{K3}}
\def\Re{\operatorname{Re}}

\def\CY{Calabi--Yau}

\def\cM{{\Scr M}}

\def\ff#1#2{{\textstyle\frac{#1}{#2}}}

\def\tick{\ding{51}}
\def\cross{\ding{55}}

\begin{document}

\begin{titlepage}
\begin{flushright}
DUKE-CGTP-05-04\\
SLAC-PUB-11257\\
SU-ITP-05/20\\
hep-th/0506014\\
May 2005
\end{flushright}
\vspace{.5cm}
\begin{center}
\baselineskip=16pt
{\fontfamily{ptm}\selectfont\bfseries\huge
Fixing All Moduli for\\[3mm] M-Theory on K3$\times$K3}\\[20mm]
{\bf\large  Paul S.~Aspinwall $^{1, 2, 3}$ and Renata Kallosh$^1$
 } \\[7mm]

{\small

$^1$ Department of Physics, Stanford University,
Stanford, CA 94305-4060\\ \vspace{6pt}
$^2$ SLAC,  Stanford, CA 94305-4060 \\ \vspace{6pt}
$^3$ Center for Geometry and Theoretical Physics, 
  Box 90318 \\ Duke University, 
 Durham, NC 27708-0318
 }

\end{center}

\begin{center}
{\bf Abstract}
\end{center}
{We analyze M-theory compactified on $\KIII\times\KIII$ with fluxes
  preserving half the supersymmetry and its F-theory limit, which is
  dual to an orientifold of the type IIB string on $\KIII\times
  (T^2/\Z_2)$. The geometry of attractive K3 surfaces plays a
  significant role in the analysis. We prove that the number of
  choices for the K3 surfaces is finite and we show how they can be
  completely classified. We list the possibilities in one case. We
  then study the instanton effects and see that they will generically
  fix all of the moduli. We also discuss situations where the
  instanton effects might not fix all the moduli.  }
\vspace{2mm} \vfill \hrule width 3.cm
\vspace{1mm}
{\footnotesize \noindent email:
psa@cgtp.duke.edu, \,  kallosh@stanford.edu}
\end{titlepage}

\vfil\break


\section{Introduction}    \label{s:intro}

One of the simplest and most accessible forms of flux compactification
is given by M-theory on $\KIII\times\KIII$. This was first a analyzed
in \cite{DRS:M-G}. The fluxes may preserve the full $N=4$
supersymmetry, or break some or all of the supersymmetry. We will be
concerned with the case where this flux breaks the supersymmetry to
$N=2$.

The F-theory limit of this theory yields an $N=1$ theory in four
dimensions and is dual, via the construction of \cite{Sen:limit1}, to
the type IIB string on $\KIII\times(T^2/\Z_2)$, where the $\Z_2$
action includes an orientifolding reflection on the world-sheet.

This theory, mainly in the orientifold language, was analyzed in
\cite{TT:K3T2}. The fluxes themselves obstruct many of the moduli of 
$\KIII\times\KIII$ but, at least if one uses the rules of the
supergravity limit described in \cite{BB:8}, one cannot fix all of the
moduli. It is believed that there are possibilities of flux
obstruction beyond those found in supergravity \cite{me:mflux}, but
the rules for this are not yet understood properly so we will not
consider this possibility in this paper.

The fluxes select a preferred complex structure on $\KIII\times\KIII$
and a given choice of flux determines this complex structure uniquely.
There remain up to 20 undetermined complexified K\"ahler moduli for
each K3 surface.  We will show that, in certain cases, all of these
remaining moduli are generically fixed by M5-brane instanton
corrections to the superpotential. 

It has been realized recently
\cite{Gorlich:2004qm,Tripathy:2005hv,Saul:flux0,KKT:flux0,BM:K3m} that
fluxes may modify Witten's \cite{Wit:super} analysis of which divisors
M5-instantons may wrap to give non-trivial effects. This allows for
interesting instanton effects even in simple geometries, such as a tori
and K3 surfaces, where na\"\i vely one might not expect such things.

In particular, in \cite{KKT:flux0} an explicit counting of fermionic
zero modes on M5 with the background (2,2) primitive flux $G$ was
performed. The generalized condition for the non-vanishing instanton
corrections to the superpotential in this case requires that the new,
flux dependent index of the Dirac operator equals to one,
$\chi_{_{D}}(G)=1$. Here $\chi_{_{D}}(G)= \chi_{_{D}}-(h^{(0,2)}-n)$
where $\chi_{_{D}}$ is the arithmetic genus of the divisor and $0 \leq
n\leq h^{(0,2)}$ is a number of solutions of a certain constraint
equation which the fermionic zero modes have to satisfy in presence of
fluxes. In absence of fluxes this condition is reduced back to
Witten's condition \cite{Wit:super} that $\chi_{_{D}}=1$.

In particular, for the case of $\KIII\times\KIII$ 4-fold and divisors
of the form $\KIII\times\P^1$ without fluxes $\chi_{_{D}}=2$ and no
instanton corrections to the superpotential are possible. In presence
of the background (2,2) primitive flux $G$ it was established in
\cite{KKT:flux0} that $n=h^{(0,2)}$, and therefore $\chi_{_{D}}(G)=1$
and instantons corrections to the superpotential are possible. The
same result for $\KIII\times\KIII$ was obtained in \cite{Saul:flux0}.

Oddly enough, we will see that if one is overzealous and tries to
leave fewer than 20 K\"ahler moduli unfixed by the flux, the
possibility arises that the instanton effects might be unable to fix
some of the remaining moduli.

Our interest in the model of M-theory compactified on
$\KIII\times\KIII$ is two-fold. First of all, this is a relatively
simple model, well-understood in the framework of IIB string theory
and 4d gauged supergravity \cite{Andrianopoli:2003jf}. The geometry of
K3 surfaces is far-better understood than generic \CY\ threefolds and
fourfolds and so this model can be analyzed more thoroughly than the
many previous examples
\cite{Denef:2004dm,Denef:2005mm,Derendinger:2005ph,DeWolfe:2005uu}
with all moduli fixed. Secondly, this model has practical applications
to cosmology of D3/D7 brane inflation in type IIB string on
$\KIII\times (T^2/\Z_2)$ \cite{Dasgupta:2002ew,Koyama:2003yc}.

The geometry of fluxes on $\KIII\times\KIII$ is a very beautiful
subject and has connections with number theory as analyzed in
\cite{Moor:attLH}. Here we will show that this allows for a complete
analysis of all possibilities. In the case that the flux is purely of
the type that breaks half the supersymmetries, we list all 13
possibilities that arise. Of these, only 8 correspond to orientifolds.

In section \ref{s:flux} we will analyze the conditions imposed on the
K3 surfaces by a flux which breaks half the supersymmetry. This
contains some very pretty mathematics associated to ``attractive K3
surfaces''. In section \ref{s:cor} we discuss the role of M5-brane
instantons and argue that all the moduli will be generically fixed,
except possibly in some cases where a particular choice of flux is
made. We conclude in section \ref{s:conc}.


\section{Moduli Spaces and Fluxes} \label{s:flux}

In this section we review the analysis of fluxes for M-theory on
$\KIII\times\KIII$ and its F-theory limit. The latter is equivalent to
an orientifold of the type IIB string on
$\KIII\times(T^2/\Z_2)$. While this has been analyzed quite
extensively in \cite{TT:K3T2}, we present a slightly different
approach which more closely follows \cite{Moor:attLH} which we believe
is a little more efficient.

\subsection{M Theory}  \label{ss:fluxM}

Let us begin with M-theory on $S_1\times S_2$, where each $S_j$ is a
K3 surface. For compactification on an 8-manifold $X$, an element of
$G$-flux may be present. This $G$-flux is subject to a quantization
condition \cite{Wit:Flux}, which asserts that, in our case\footnote{We
  have absorbed a factor of $2\pi$ into $G$ compared to much of the rest
  of the literature.}
\begin{equation}
  G \in H^4(S_1\times S_2,\Z).
\end{equation}
A consistent theory must contain M2-branes and/or nonzero
$G$-flux in this background satisfying \cite{SVW:8}
\begin{equation}
  n_{\textrm{M2}} + \ff12 G^2 = 24.  \label{eq:tadpole}
    \end{equation}
The M2-branes will not break any supersymmetry, but the $G$-flux
may. The supergravity analysis of \cite{BB:8} showed that $G$ must be
{\em primitive\/} and of type (2,2) in order that any supersymmetry be
preserved. Any such integral 4-form may be decomposed
\begin{equation}
  G = G_0 + G_1, \label{eq:Ggen}
\end{equation}
where
\begin{equation}
\begin{split}
 G_0 &= \sum_{\alpha=1}^M \omega_1^{(\alpha)}\wedge\omega_2^{(\alpha)}\\
 G_1 &= \Re(\gamma\Omega_1\wedge\overline{\Omega}_2), 
\end{split}  \label{eq:G0G1}
\end{equation}
and $\omega_j^{(\alpha)}$ are (cohomology classes\footnote{We only
  discuss {\em cohomology classes\/} of forms in this paper but we
  will usually not state this explicitly to avoid cluttering
  notation.} of) integral primitive (1,1)-forms on $S_j$, $\Omega_j$
  is the holomorphic 2-form on $S_j$ and $\gamma$ is a complex number
  which must be chosen to make the last term integral.

There are essentially three kinds of moduli which arise in such a
compactification:
\begin{enumerate}
\item Deformations of the K3 surfaces $S_1$ and $S_2$.
\item Motion of the M2-branes.
\item Deformations of vector bundles with nonabelian structure group
  associated to enhanced gauge symmetries arising from singular points
  in $S_1$ and $S_2$.
\end{enumerate}
By assuming, from now on, that $n_{M2}=0$ and that our K3 surfaces are
smooth, we will restrict attention to only the first kind of modulus
in this paper.

The moduli space of M-theory on $\KIII\times\KIII$ is of the form
$\cM_1\times\cM_2$, where each factor is associated to one of the K3
surfaces. If no supersymmetry is broken by fluxes, each of the $\cM_j$
factors is a quaternionic K\"ahler manifold. Ignoring instanton
corrections, each $\cM_j$ is of the form
\begin{equation}
  \GO(\Gamma_{4,n_j})\backslash\GO(4,n_j)/(\GO(4)\times\GO(n_j)),
    \quad j=1,2
       \label{eq:M1}
\end{equation}
where $\Gamma_{4,n_j}$ is a lattice of signature $(4,n_j)$. The values
of $n_j\leq20$ will be determined by the choice of flux $G$. The space
(\ref{eq:M1}) should be viewed as the Grassmannian of space-like
4-planes in $\Pi_j\subset\Gamma_{4,n_j}\otimes\R$ divided out by the
discrete group of automorphisms of the lattice $\Gamma_{4,n_j}$.

The Grassmannian (\ref{eq:M1}) is familiar (for $n_j=20$) from the
moduli space of $N=(4,4)$ superconformal field theories associated to
the sigma model with a K3 target space $S$. In this case, the degrees
of freedom parametrized by the conformal field theory are given by a
Ricci flat metric on $S$ together with a choice of $B\in
H^2(S,\GU(1))$. The choice of metric on a K3 surface of volume one is
given by a space-like 3-plane $\Sigma\subset H^2(S,\R)=\R^{3,19}$. The
3-plane $\Sigma$ is spanned by the real and imaginary parts of
$\Omega$, and the K\"ahler form $J$.  The extra data of the $B$-field
and volume extend this to a choice of space-like 4-plane $\Pi\subset
H^*(S,\R)=\R^{4,20}$. We refer to \cite{me:lK3} and references therein
for a full account of this.

Even though M-theory itself has no $B$-field, the M5-brane wrapped on
one $\KIII$ gives us an effective $B$-field for compactification on
the other K3. Hence the form (\ref{eq:M1}). We refer to
\cite{me:mflux} for examples.

A K3 surface is a hyperk\"ahler manifold and thus has a choice of
complex structures for a fixed metric. This choice corresponds to
specifying the direction of $J$ in $\Sigma$. Since supersymmetries are
constructed for complex structures, this multiplicity of complex
structures implies the existence of a specific extended supersymmetry.

If $G_1=0$ in (\ref{eq:Ggen}) then the condition that
$G$ be primitive and of type (2,2) preserves the freedom to rotate
$\Omega$ and $J$ within $\Sigma$. Thus, values of $G$ purely of the
form $G_0$ preserve the full $N=4$ supersymmetry in three dimensions
\cite{DRS:M-G}.

If the term $G_1$ in (\ref{eq:Ggen}) is non-trivial then we destroy the
symmetry of rotations within $\Sigma$ and the supersymmetry is broken
to $N=2$. This is the case of interest and we therefore assume, from now
on, that $G_1$ is nonzero.

\subsection{Attractive K3 surfaces}   \label{ss:attr}

For now, let us assume that $G$ is purely of type $G_1$, i.e.,
$G_0=0$.  Let
\begin{equation}
\Omega_j = \alpha_j+i\beta_j.
\end{equation}
for $\alpha_j,\beta_j\in H^2(S_j,\R)$. 
From $\int_{S_j} \Omega_j\wedge\overline{\Omega}_j>0$ and
$\int_{S_j} \Omega_j\wedge{\Omega}_j=0$, 
it follows that\footnote{We use the implicit inner product $a.b=\int_S
  a\wedge b$.}
\begin{equation}
\begin{split}
\alpha_j^2 &= \beta_j^2 > 0\\
\alpha_j.\beta_j &= 0\\
\alpha_j&\neq\beta_j
\end{split}  \label{eq:Omega}
\end{equation}

We also have
\begin{equation}
G = \alpha_1\wedge\alpha_2 + \beta_1\wedge\beta_2, \label{eq:Gab}
\end{equation}
where we set $\gamma=1$ in (\ref{eq:G0G1}) by rescaling $\Omega_1$.
Let $\mathbf{\Omega}_j$ be a 2-plane in $H^2(S_j,\R)$ spanned by
$\alpha_j$ and $\beta_j$. We claim
\begin{theorem}
$\mathbf{\Omega}_1$ and $\mathbf{\Omega}_2$ are uniquely determined by
  $G$.  \label{th:fix}
\end{theorem}
To prove this we first use the K\"unneth formula which tells us that
\begin{equation}
H^4(S_1\times S_2,\Z) \cong H^0(S_1,\Z)\otimes H^4(S_2,\Z) \oplus
H^2(S_1,\Z)\otimes H^2(S_2,\Z) \oplus H^4(S_1,\Z)\otimes H^0(S_2,\Z).
     \label{eq:Kun}
\end{equation}
We know from (\ref{eq:G0G1}) that $G$ lies entirely in the second term
on the right-hand side of (\ref{eq:Kun}). Let us assume we are given
$G,\alpha_j,\beta_j$ solving
\begin{equation}
  G = \alpha_1\otimes\alpha_2 +
      \beta_1\otimes\beta_2.
\end{equation}
Now try to find other solutions of the form
\begin{equation}
  G = (\alpha_1+\alpha_1')\otimes(\alpha_2+\alpha_2') +
      (\beta_1+\beta_1')\otimes(\beta_2+\beta_2').
\end{equation}
It follows that
\begin{equation}
  \alpha_1\otimes\alpha_2'+\alpha_1'\otimes\alpha_2+
  \alpha_1'\otimes\alpha_2' +
  \beta_1\otimes\beta_2'+\beta_1'\otimes\beta_2+
  \beta_1'\otimes\beta_2' = 0.
\end{equation}
Let $\pi_1$ be the projection
\begin{equation}
\pi_1:H^2(S_1,\R)\to
H^2(S_1,\R)/\Span(\alpha_1,\beta_1).
\end{equation}
Thus
\begin{equation}
\pi_1(\alpha_1')\otimes(\alpha_2+\alpha_2')+
\pi_1(\beta_1')\otimes(\beta_2+\beta_2')=0.
\end{equation}
The only solution is to put $\pi_1(\alpha_1')=\pi_1(\beta_1')=0$,
which corresponds to not rotating $\mathbf{\Omega}_1$ at all; or
putting $\alpha_2+\alpha_2'$ or $\beta_2+\beta_2'$ equal to zero, or
making $\alpha_2+\alpha_2'$ and $\beta_2+\beta_2'$ collinear. The latter
conditions would make the new $\mathbf{\Omega}_2$, spanned by
$\alpha_1+\alpha_1'$ and $\beta_1+\beta_1'$ violate
(\ref{eq:Omega}). We may also reverse the r\^oles of
$\mathbf{\Omega}_1$ and $\mathbf{\Omega}_2$ in the argument. This
completes the proof of theorem \ref{th:fix}.

\vspace{3mm}

The statement that $\mathbf{\Omega}_1$ and $\mathbf{\Omega}_2$ are
fixed by $G$ means that {\em the complex structures of $S_1$ and $S_2$
  and uniquely determined by a choice of flux}.

\vspace{3mm}

The next thing we prove is
\begin{theorem}
The K3 surfaces $S_1$ and $S_2$ whose complex structures are fixed by
$G$ are forced to both be attractive. \label{th:att}
\end{theorem}
Before we prove this, we first review the definition of an
    attractive\footnote{The standard mathematical term is ``singular''
    but as this is such a singularly misleading term, we prefer to
    follow Moore's choice of language from \cite{Moor:attLH}.}  
K3 surface.  The {\em Picard lattice\/} of a K3 surface is given by
the lattice $H^{1,1}(S_j)\cap H^2(S_j,\Z)$. The {\em Picard
number\/} $\rho(S_j)$ is defined as the rank of this lattice. The
surface $S_j$ is said to be {\em attractive\/} if $\rho(S_j)=20$,
the maximal value.

Let us define
\begin{equation}
\Upsilon_j = \Bigl(H^{2,0}(S_j)\oplus H^{0,2}(S_j)\Bigr)\cap H^2(S_j,\Z),
\end{equation}
which is the intersection of the 2-plane $\mathbf{\Omega}_j$ with the
lattice $H^2(S_j,\Z)$ in the space $H^2(S_j,\R)$. For a generic K3
surface $\Upsilon_j$ will be completely trivial, but the maximal rank
of $\Upsilon_j$ is 2. The ``transcendental lattice'' is defined as the
orthogonal complement of the Picard lattice in $H^2(S_j,\Z)$. If, and
only if, the rank of $\Upsilon_j$ is 2, the transcendental lattice
will coincide with $\Upsilon_j$ and the K3 surface $S_j$ will be
attractive.  We therefore need to prove that $\Upsilon_j$ is rank 2.
 
Let $e^j_k$, $k=1,\ldots,22$ be an integral basis for
$H^2(S_j,\Z)$. Expanding
\begin{equation}
\begin{split}
\alpha_j &= \sum_k a_{jk}e^j_k\\
\beta_j &= \sum_k b_{jk}e^j_k\\
G &= \sum_{kl} N_{kl}e^1_k\otimes e^2_l,
\end{split}
\end{equation}
where $a_{jk}$ and $b_{jk}$ are real numbers and $N_{kl}$ are integers
(since $G$ is an {\em integral\/} 4-form).  Then (\ref{eq:Gab})
becomes
\begin{equation}
  a_{1k}a_{2l} + b_{1k}b_{2l} = N_{kl},\quad\hbox{for all $k,l$}.
\end{equation}
Fixing $l$, the above equation may be read as saying that a real
combination of $\alpha_1$ and $\beta_1$ lies on a lattice point of
$H^2(S_1,\Z)$. By varying $l$ we get 22 different such combinations. The
fact that $\alpha_2$ and $\beta_2$ are linearly independent means that
all these lattice points cannot be collinear. Thus $\mathbf{\Omega}_1$
contains a 2-dimensional lattice. Similarly $\mathbf{\Omega}_2$
contains a 2-dimensional lattice and we complete the proof of theorem
\ref{th:att}.

\vspace{3mm}

Attractive K3 surfaces were completely classified in
\cite{SI:singK3}. They were shown to be in one-to-one correspondence
with $\Sl(2,\Z)$-equivalence classes of positive-definite even
integral binary quadratic forms. Such a quadratic form can be written
in terms of a matrix
\begin{equation}
  Q = \begin{pmatrix}2a&b\\b&2c\end{pmatrix}, \label{eq:Q1}
\end{equation}
where $a,b,c\in\Z$, $a>0$, $c>0$, and $\det Q = 4ac-b^2>0$. Two forms
$Q$ and $Q'$ define an equivalent K3 surface if, and only if,
$Q'=M^TQM$, for some $M\in\Sl(2,\Z)$.

Let $\Upsilon_j$ be spanned (over the integers) by integral vectors
$p_j$ and $q_j$. The above lattice is then
\begin{equation}
  Q_j = \begin{pmatrix}p_j^2&p_j.q_j\\
    p_j.q_j&q_j^2\end{pmatrix},
\end{equation}
We are free to rescale $\Omega_1$ and $\Omega_2$ (since they are only
defined up to complex multiplication) so that
\begin{equation}
  \Omega_j = p_j + \tau_j q_j,
\end{equation}
for a complex number $\tau_j$, which is fixed by the condition
$\Omega_j^2=0$ to be
\begin{equation}
  \tau_j = \frac{-p_j.q_j + i\sqrt{\det Q_j}}{q_j^2}.
\end{equation}
Note that this choice of rescaling means we cannot now assume
$\gamma=1$ in (\ref{eq:G0G1}).
We then obtain
\begin{equation}
  G = \Bigl(\Re(\gamma)p_1\otimes p_2 + \Re(\gamma\tau_1)q_1\otimes p_2 +
   \Re(\gamma\bar\tau_2)q_2\otimes p_1 + \Re(\gamma\tau_1\bar\tau_2) q_1\otimes
   q_2\Bigr). \label{eq:Gint}
\end{equation}
Consider the condition imposed by the integrality of $G$.  Since
$p_1\otimes p_2$ is integral and primitive\footnote{Primitive in the
sense that it is not an integral multiple of a lattice element.}  we
must have $\Re(\gamma)\in\Z$. The other terms on (\ref{eq:Gint}) put
further conditions of $\gamma$. It is easy to show that a consistent
choice of $\gamma$ making each term in (\ref{eq:Gint}) integral is
possible if and only if $\sqrt{\det(Q_1Q_2)}$ is an integer. That is,
\begin{theorem}
A pair of attractive K3 surfaces $S_1$ and $S_2$ will correspond to a
choice of integral $G$-flux if and only if $\det(Q_1Q_2)$ is a perfect
square.
\end{theorem}

Finally we need to impose the tadpole condition $\ff12 G^2=24$.
We compute
\begin{equation}
G^{2}=\ff14(\gamma\Omega_1\wedge\overline{\Omega}_2+
\bar\gamma\Omega_2\wedge\overline{\Omega}_1)^{2},
\end{equation}
and use the fact that $\Omega_{j}^{2}=0$ so only the cross term
in the square is not vanishing. Therefore
\begin{equation}
G^{2}= \ff12|\gamma|^2\int \Omega_1\wedge\overline{\Omega}_1 
   \int \Omega_2\wedge\overline{\Omega}_2.
\end{equation}

Using  $ \Omega_j = p_j + \tau_j q_j$ we find
\begin{equation}
\ff12G^{2}= \frac{|\gamma|^2\det(Q_{1} Q_{2})}{q_{1}^{2 }q_{2}^{2}}=24.
\label{eq:tad1}
\end{equation}

Solving (\ref{eq:tad1}) together with the integrality of
(\ref{eq:Gint}) provides all possibilities of flux compactifications
with $G=G_1$. We may prove that there is a finite number of attractive
K3 surfaces that yield solutions to this equation as follows. We use
the following theorem from \cite{Keng:NT}:
\begin{theorem}
In the equivalence class of the matrix (\ref{eq:Q1}) under the action
of $\Sl(2,\Z)$, assuming that $-\det(Q)$ is not a perfect square, one
can always find a representative matrix satisfying
\begin{equation}
  |b| \leq |c| \leq |a|.
\end{equation}   \label{th:bound}
\end{theorem}
In our case $-\det(Q)$ is negative and so, clearly, not a perfect
square. Thus we may restrict attention to matrices satisfying the
above bounds.  Putting
\begin{equation}
  Q_1 = \begin{pmatrix}2a&b\\b&2c\end{pmatrix}, \qquad
  Q_2 = \begin{pmatrix}2d&e\\e&2f\end{pmatrix},
\end{equation}
yields
\begin{equation}
\det(Q_1) = 4ac-b^2 \geq 4ac-ac = 3ac.
\end{equation}
Similarly, $\det(Q_2)\geq 3df$. Thus (\ref{eq:tad1}) yields
\begin{equation}
\ff12 G^2 \geq \ff94 |\gamma|^2 ad.
\end{equation}
Suppose $\Re(\gamma)\neq0$. Then $|\gamma|^2\geq1$, since
$\Re(\gamma)\in\Z$. In this case we are done since $a$ and $d$ are
positive integers and $b,c,e,f$ are constrained by theorem
\ref{th:bound}. On the other hand, if $\gamma$ is purely imaginary,
then the integrality of the second term in (\ref{eq:Gint}) forces
\begin{equation}
  \frac{\Im(\gamma)\sqrt{\det(Q_1)}}{2c}\in\Z.
\end{equation}
Obviously $\Im(\gamma)$ cannot be zero since then $|\gamma|^2$ would
be zero. We then have
\begin{equation}
\ff12 G^2 \geq \frac{c\det(Q_2)}{f}\geq 3dc.
\end{equation}
This bounds $c$ and $d$. Similarly we may use the third term in
(\ref{eq:Gint}) to bound $a$ and $f$. Thus we complete the proof that
there are only a finite number of possibilities for $a,b,c,d,e,f$, and
thus only a finite number of attractive K3 surfaces whenever $G^2$ is
bounded.

In fact, it is not hard to perform a computer search to yield the full
list of possibilities. For $\ff12G^2=24$, i.e., $G_0=0$, there are 13
possibilities up to $\Sl(2,\Z)$ equivalence which we list in table
\ref{tab:l}. The column labeled ``O?'' will be explained in section
\ref{ss:orient}. In principle, a given pair of attractive K3 surfaces
might admit many, but finitely many, inequivalent choices of $G$. In
our case, where the numbers are quite small, this never happens.

One may, of course, obtain other possibilities by considering a
nonzero $G_0$. In this case, we solve the same problem for
$\ff12G^2<24$. As one might expect, the number of possibilities for a
given $\ff12G^2<24$ are somewhat fewer than above.

\begin{table}
\[
\begin{array}{|cc|c|c||cc|c|c|}
\hline
Q_1&Q_2&\gamma&\hbox{O?}&Q_1&Q_2&\gamma&\hbox{O?}\\
\hline
&&&&&&&\\[-3mm]
\begin{pmatrix}4&0\\0&2\end{pmatrix}&
\begin{pmatrix}4&0\\0&2\end{pmatrix}&
1+\ff i{\sqrt{2}}&\hbox{\cross}&
\begin{pmatrix}4&2\\2&4\end{pmatrix}&
\begin{pmatrix}2&1\\1&2\end{pmatrix}&
2+\ff{2i}{\sqrt{3}}&\hbox{\tick}\\[4mm]
\begin{pmatrix}4&2\\2&4\end{pmatrix}&
\begin{pmatrix}6&0\\0&2\end{pmatrix}&
1+\ff{i}{\sqrt{3}}&\hbox{\tick}&
\begin{pmatrix}6&0\\0&4\end{pmatrix}&
\begin{pmatrix}6&0\\0&4\end{pmatrix}&
\ff{2i}{\sqrt{6}}&\hbox{\cross}\\[4mm]
\begin{pmatrix}6&0\\0&6\end{pmatrix}&
\begin{pmatrix}2&0\\0&2\end{pmatrix}&
1+i&\hbox{\cross}&
\begin{pmatrix}6&0\\0&6\end{pmatrix}&
\begin{pmatrix}4&0\\0&4\end{pmatrix}&
1&\hbox{\tick}\\[4mm]
\begin{pmatrix}8&4\\4&8\end{pmatrix}&
\begin{pmatrix}4&2\\2&4\end{pmatrix}&
1+\ff{i}{\sqrt{3}}&\hbox{\tick}&
\begin{pmatrix}12&0\\0&2\end{pmatrix}&
\begin{pmatrix}12&0\\0&2\end{pmatrix}&
\ff{i}{\sqrt{6}}&\hbox{\cross}\\[4mm]
\begin{pmatrix}12&0\\0&4\end{pmatrix}&
\begin{pmatrix}2&1\\1&2\end{pmatrix}&
1+\ff{i}{\sqrt{3}}&\hbox{\tick}&
\begin{pmatrix}12&0\\0&4\end{pmatrix}&
\begin{pmatrix}6&0\\0&2\end{pmatrix}&
\ff{i}{\sqrt{3}}&\hbox{\tick}\\[4mm]
\begin{pmatrix}12&0\\0&6\end{pmatrix}&
\begin{pmatrix}4&0\\0&2\end{pmatrix}&
\ff{i}{\sqrt{2}}&\hbox{\cross}&
\begin{pmatrix}12&0\\0&12\end{pmatrix}&
\begin{pmatrix}2&0\\0&2\end{pmatrix}&
1&\hbox{\tick}\\[4mm]
\begin{pmatrix}16&8\\8&16\end{pmatrix}&
\begin{pmatrix}2&1\\1&2\end{pmatrix}&
1+\ff{i}{\sqrt{3}}&\hbox{\tick}&&&&\\[4mm]
\hline
\end{array}
\]
\caption{The 13 pairs of matrices $Q_1,Q_2$ yielding the possible
  attractive K3 surfaces. The column headed
``O?'' shows 8 solutions when one of K3 surfaces  is a ``Kummer
surface''.} \label{tab:l}
\end{table}

As stated above, the complex structures of $S_1$ and $S_2$ are
fixed. What remains unfixed is the K\"ahler form and $B$-field degree
of freedom. Using the assumption $G_0=0$ in (\ref{eq:Ggen}), all
20 such complex degrees of freedom remain undetermined by the
fluxes. A non-trivial choice of $G_0$ will fix some of these 20
remaining moduli.

The choice of $G$ fixes a 2-plane within $\Pi_j$ spanned by the real
and imaginary parts of $\Omega_j$. This means that the moduli space
(\ref{eq:M1}) is reduced to
\begin{equation}
  \cM_j\cong\GO(\Gamma_{2,n_j})\backslash\GO(2,n_j)/(\GO(2)\times\GO(n_j)).
       \label{eq:M2}
\end{equation}
If $G_0=0$ then $n_1=n_2=20$. If $G_0$ is nonzero, these numbers will decrease.

\subsection{The Orientifold}  \label{ss:orient}

Now let us turn our attention to the related question of orientifolds
on $\KIII\times(T^2/\Z_2)$. One obtains this orientifold via F-theory.

Begin with M-theory on $S_1\times S_2$ (ignoring flux for now) to obtain
an $N=4$ theory in three dimensions as above. Assume that $S_2$ is an
elliptic K3 surface with a section. Let $\pi:S_2\to B$ denote this
elliptic fibration of $S_2$. By shrinking the area of the elliptic
fibre, one moves to an F-theory fibration corresponding to a type IIB
compactification on $S_1\times B$. This yields a four-dimensional
$N=2$ compactification.

This four-dimensional theory can be compactified on a circle thus
regaining the three-dimensional theory we had originally from the
M-theory compactification. The relationship between the moduli spaces
of the three-dimensional theory and four-dimensional theory can be
understood from this fact. The moduli space of the three-dimensional
theory is $\cM_1\times \cM_2$, where each $\cM_j$ is quaternionic
K\"ahler. The four-dimensional theory has a moduli space
$\cM_H\times\cM_V$, where $\cM_H$, the hypermultiplet moduli space, is
exactly $\cM_1$.

The vector multiplet moduli space $\cM_V$, is special K\"ahler. The
complex dimension of $\cM_V$ is one less than the quaternionic
dimension of $\cM_2$. Quantum corrections make for a very complicated
relationship between $\cM_V$ and $\cM_2$. Let us ignore these quantum
corrections for now, which we may do since we are only making
qualitative statements about the moduli space. In this case, ignoring
any flux effects or M2-branes, we have, locally
\begin{equation}
  \cM_2 = \frac{\GO(4,20)}{\GO(4)\times\GO(20)},
\end{equation}
and, from the $c$-map \cite{CFG:II}
\begin{equation}
  \cM_V = \frac{\GO(2,18)}{\GO(2)\times\GO(18)} \times
  \frac{\Sl(2,\R)}{\GU(1)}. \label{eq:Mv}
\end{equation}
The first factor of (\ref{eq:Mv}) corresponds to the complex structure
moduli space of $S_2$ if we declare $S_2$ to be an elliptic fibration
with a section. In F-theory language, this corresponds to the moduli
space of the location of 7-branes. The second factor of (\ref{eq:Mv})
would na\"\i vely correspond to the complexified area of the base, $B$, of
the elliptic fibration as this is the only modulus remaining once the
fibre is shrunk to zero size. When moving between dimensions one must
be careful with taking into account overall scalings of the
metric. The result is that the second factor of (\ref{eq:Mv}) actually
corresponds to the complexified volume of the K3 surface $S_1$.  The
area of the base becomes a parameter in the hypermultiplet moduli
space $\cM_H$. We refer to \cite{Andrianopoli:2003jf} for more details.

Sen \cite{Sen:limit1} showed how type IIB orientifolds could be
obtained from F-theory compactifications. Elliptic fibrations may
contain ``bad fibres'', i.e., fibres which are not elliptic
curves. These bad fibres have been classified by Kodaira. We refer to
\cite{me:lK3}, for example, for a review.  In Sen's analysis one takes
a limit in the moduli space of complex structures of the elliptic
fibration such that all the bad fibres become type $\mathrm{I}_0^*$ in
the Kodaira classification. We now have a type IIB string compactified
on the orientifold $S_1\times (C/\Z_2)$, where $C$ is an elliptic
curve and the base of the elliptic fibration is $B\cong C/\Z_2$.

This limit freezes the location of the F-theory
7-branes making the moduli space locally
\begin{equation}
  \cM_V = \frac{\Sl(2,\R)}{\GU(1)}\times\frac{\Sl(2,\R)}{\GU(1)}\times
  \frac{\Sl(2,\R)}{\GU(1)}.  \label{eq:Mv-or}
\end{equation}
These three complex moduli can be identified as
\begin{itemize}
\item The modulus of the F-theory elliptic fibre --- i.e., the
  axion-dilaton of the type IIB string.
\item The modulus of the elliptic curve $C$, where the base of the
  elliptic fibration of $S_2$ is $B\cong C/\Z_2$.
\item The complexified volume of $S_1$, as above.
\end{itemize}
The 16 moduli that we have ``lost'' in passing from (\ref{eq:Mv}) to
(\ref{eq:Mv-or}) are regained by allowing D7-branes to move away from
the 4 O7-planes.

\vspace{5mm}

Now consider the effect of flux in the form of $G_1$ so as to yield an
$N=1$ supersymmetric theory in four dimensions. This flux fixes the
complex structure of $S_1$ and $S_2$ making both of these K3 surfaces
attractive. First note that any attractive K3 surface is elliptic
with a section \cite{SI:singK3} so our condition for an F-theory limit
is automatically satisfied.

The flux causes the dimension of $\cM_H\cong\cM_1$ to be halved ---
exactly as it was in the case of M-theory in section
\ref{ss:attr}. For $\cM_V$, the first factor of (\ref{eq:Mv}) is a
complex structure moduli space and so disappears completely. In
orientifold language, we fix the dilaton-axion, the complex structure
of $C\cong T^2$, and the location of all the D7-branes.  All that
remains unfixed in $\cM_V$ is a single complex modulus corresponding
to the complexified volume of $S_1$.

Sen's orientifold limit of F-theory is a limit of complex structure,
but once we turn on flux, we have no deformations of the complex
structure! The only way our M-theory compactification can correspond
to an orientifold is if the elliptic fibration of the attractive $S_2$
has this fibration structure to begin with.

So let us suppose $S_2$ is an attractive K3 surface which is an
elliptic fibration with only smooth, or type $\mathrm{I}_0^*$
fibres. The base of such a fibration must be $\P^1$ and there must be
exactly four $\mathrm{I}_0^*$ fibres and no other singular fibres. In
this case, the $J$-invariant of the fibre has no zeros or poles and
is therefore a constant. This is exactly the same elliptic fibration
data as one would obtain for a K3 surface which is a ``Kummer
surface'', i.e., a blow-up of a quotient $A/\Z_2$, where $A$ is a
4-torus (or abelian surface to be more precise). It follows that $S_2$
is indeed a Kummer surface (following, for example, proposition 2.7 of
\cite{Mir:mod}).

Any Kummer surface, which is attractive, must be a
$\Z_2$-quotient of an attractive abelian surface (see, for example,
equation (5.8) of \cite{SM:abel}). Such abelian surfaces are
classified in much the same way as attractive K3 surfaces, that is,
they are again in one-to-one correspondence with
$\Sl(2,\Z)$-equivalence classes of positive-definite even integral
binary quadratic forms. If $Q$ is the matrix associated with the
binary quadratic form of our attractive Kummer surface $S_2$ and $R$ is
the matrix associated with the attractive abelian surface $A$, then
one can show that (see \cite{BPV:}, for example)
\begin{equation}
  Q = 2R.
\end{equation}
It follows that the attractive K3 surface $S_2$ is a Kummer surface
if, and only if, the associated even binary quadratic form is twice
another even binary quadratic form. {\em Only the F-theory
compactifications on $\KIII\times\KIII$ which satisfy this property
will have orientifold interpretations.} 

Looking back at table \ref{tab:l}, we see that 8 of our 13
possibilities admit an orientifold interpretation. The column headed
``O?'' denotes whether an orientifold model exists.

One might be concerned that one should check that $G$ is compatible
with the F-theory limit as spelt out in \cite{GVW:8d}. That is, $G$
should have ``one leg'' in the fibre direction. This condition turns
out to be automatically satisfied, at least in the case $G_0=0$, as we
explain as follows.

The spectral sequence for the cohomology of a fibration yields
\begin{equation}
  H^2(S,\R) = H^0(B,R^2\pi_*\R) \oplus H^1(B,R^1\pi_*\R) \oplus 
    H^2(B,\pi_*\R),  \label{eq:spec}
\end{equation}
where $H^p(B,R^q\pi_*\R)$ may be schematically viewed as a form with 
$p$ legs in the base direction and $q$ legs in the fibre direction.

The term $H^2(B,\pi_*\R)$ is dual to the base $B\cong\P^1$. The term
$H^0(B,R^2\pi_*\R)$ is dual to the fibres including components of
singular fibres. Both of these terms correspond to curves in $S$ and
thus forms of type (1,1). Therefore any form of type (0,2) or (2,0)
must be contained $H^1(B,R^1\pi_*\R)$. It follows that $G$ has one leg
in the fibre direction assuming $G=G_1$ in (\ref{eq:Ggen}).

\vspace{3mm}

Any attractive abelian surface $A$ must be of the form $C\times C'$,
where $C$ and $C'$ are isogenous elliptic curves admitting complex
multiplication \cite{Mum:Abel}. Here, ``isogenous'' means that $C'$ is
isomorphic, as an elliptic curve, to a free quotient of $C$ by any
finite subgroup of $\GU(1)\times\GU(1)$. We refer to
\cite{GV:rat,Moor:attLH} and references therein for a nice account of
complex multiplication. 

The elliptic fibration of the Kummer surface $S_2$ will therefore be
an elliptic fibration with base $C/\Z_2$ with fibre $C'$. It follows
from Sen's argument \cite{Sen:ort} that {\em F-theory on $S_1\times S_2$ is
equivalent to the type IIB orientifold on $S_1\times (C/\Z_2)$ where
the dilaton-axion of the type IIB theory is given by the
$\tau$-parameter of the elliptic curve $C'$.}

The fixing of the complex structures of $C$
and $C'$ account for the removal of the first two factors of
(\ref{eq:Mv-or}) in the vector multiplet moduli space. The fact that
$C$ and $C'$ are isogenous means that their $\tau$-parameters will be
related by an $\Gl(2,\Q)$ transformation. In other words,
\begin{equation}
  \tau_{C'} = \frac{a\tau_C +b}{c\tau_C +d},
\end{equation}
where $a,b,c,d$ are integers {\em not\/} necessarily satisfying
$ad-bc=1$.%
\footnote{There is an example in \cite{TT:K3T2} which appears to
  violate this condition. This is because the basis defined in the
  appendix of \cite{TT:K3T2} is not a valid integral basis for
  $H^2(S,\Z)$ and so the resulting $G$ is not actually in integral
  cohomology.}

We should note the fact that an attractive abelian surface may, in
general, be decomposed into $C\times C'$ in many inequivalent ways
(other than the trivial exchange of $C$ and $C'$). Thus, a fixed
$S_1\times S_2$ might be associated to none, or many inequivalent
orientifold limits. An algorithm for determining a complete set of
such factorizations was presented in \cite{SM:abel}. For example, if
the abelian surface corresponds to
\begin{equation}
  R = \begin{pmatrix}12&6\\6&12\end{pmatrix},
\end{equation}
then one may factorize into a pair of elliptic curves with
$\tau_C=\omega$ and $\tau_{C'}=6\omega$; or $\tau_C=2\omega$ and
$\tau_{C'}=3\omega$, where $\omega=\exp(2\pi i/3)$. In our cases,
listed in table \ref{tab:l}, such an ambiguity never occurs.


\section{Instanton Corrections}  \label{s:cor}

So far we have completely ignored any quantum corrections to the
moduli space. Consider first the case of M-theory on $S_1\times S_2$
where the flux does not break any supersymmetry. This yields an $N=4$
theory in three dimensions. By the usual counting, any instanton
solution that breaks half the supersymmetry will modify the
prepotential and thus deform the metric on the moduli space. These
instantons will not obstruct any moduli and the dimension of the
moduli space will be unchanged by these quantum corrections.

The only source of such instanton corrections in M-theory will
correspond to M5-brane instantons wrapping holomorphically embedded
complex 3-folds within $S_1\times S_2$. Such divisors are clearly of
the form $S_1\times C_g$, or $C_g\times S_2$, where $C_g$ is an
algebraic curve of genus $g$.

Following, \cite{Wit:super}, one can show that these divisors will
only contribute nontrivially to the prepotential if they have
holomorphic Euler characteristic $\chi_{\O}=2$. Since
$\chi_{\O}(\KIII\times C_g)=2(1-g)$, we see that our instantons must
be of the form $S_1\times\P^1$ or $\P^1\times S_2$.

Now suppose we turn flux on so as to break half the supersymmetry. 
The superpotential of the resulting low-energy effective theory will
now receive instanton corrections from M5-branes wrapping divisors. A
na\"\i ve interpretation of \cite{Wit:super} would lead one to believe
that one would look for divisors with $\chi_{\O}=1$. There are no such
divisors in $S_1\times S_2$ and so one would arrive at the conclusion
that the K\"ahler moduli cannot be removed.

This is not the case however. It was shown in
\cite{Saul:flux0,KKT:flux0,BM:K3m} that some fermion zero modes on the
M5-brane worldvolume are lost changing the counting argument of
\cite{Wit:super}. The result is that, with the $G$-flux we are using,
the desired instantons should have $\chi_{\O}=2$. That is, the
instantons which contribute to the superpotential are precisely those
wrapping $S_1\times\P^1$ or $\P^1\times S_2$.

As discussed in the previous section, the complex structure on $S_1$
and $S_2$ is completely fixed by the choice of $G$-flux. Each K3
surface is attractive and, as such has Picard number equal to
20. This leaves each K3 surface with 20 complexified K\"ahler form
moduli. If $G$ is purely of the form
$G=G_1=\Re(\Omega_1\wedge\overline{\Omega}_2)$, then these 20 moduli
are unfixed by the fluxes. Any terms from $G_0$ in (\ref{eq:Ggen}) will
fix some of these remaining 20 moduli.

In any case, at least in the supergravity approximation, one cannot
remove all of these K\"ahler moduli by fluxes. It is possible to fix
at least 10 of the K\"ahler moduli but in the F-theory limit one is
restricted to fixing only 2 K\"ahler moduli using $G_0$ effects.  It
is conceivable that going beyond the supergravity approximation may
change such statements as discussed in \cite{me:mflux}.

Let $S$ be an attractive K3 surface and let
$V=\Pic(S)\otimes\R=\R^{20}$ be the subspace of $H^2(S,\R)$ spanned by
the K\"ahler form.  We wish to find a convenient basis for $V$. Let us
consider an element of $H^2(S,\R)$ as a homomorphism from 2-chains in
$S$ to $\R$. If $\alpha\in H^2(S,\R)$ and $x$ is a 2-chain, we thus
denote $\alpha(x)\in\R$.

The following proposition will be useful
\begin{prop}
We may find a set $\{e_1,\ldots,e_{20}\}$ of holomorphically embedded
$\P^1$'s in $S$ and a basis $\{\xi_1,\ldots,\xi_{20}\}$ of $V$ such
that $\xi_a(e_b)=\delta_{ab}$. \label{prop:1}
\end{prop}
To see this we use the fact that any attractive $S$ is an elliptic
fibration $\pi:S\to B$ with at least one section as noted in section
\ref{ss:orient}. Now take any rational curve (i.e., holomorphically
embedded $\P^1$) $C\subset S$. Since $\pi$ is a holomorphic map, the
image of $C$ under $\pi$ is either a point or all of $B$. In the
former case $C$ is a component of a singular fibre and in the latter
case $C$ is a ``section'' (or multisection) of the fibration.

In theorem 1.1 of \cite{Shio:ell} it is shown that the complete Picard
lattice of an elliptic surface with a section is generated by rational
combinations of sections, smooth fibres and components of singular
fibres. If there is at least one bad fibre which is reducible, the
smooth elliptic fibre itself is homologous to a sum of smooth rational
curves. This is indeed the case for attractive K3 surfaces as shown in
\cite{SI:singK3}. The proposition then follows.

\vspace{3mm}

An instanton correction to the superpotential from an M5-brane wrapping
a divisor $D$ will be of the form $\sim f\exp(-\Vol(D))$, where
$\Vol(D)$ is the complexified volume of $D$. The coefficient $f$
may depend on complex structure moduli but cannot depend on the
K\"ahler moduli. This is because $f$ is computed perturbatively and
the ``axionic'' shift symmetry of the complex partner to the K\"ahler
form prevents any contribution to perturbation theory.

Using the bases $\{e_1^{(1)},\ldots,e_{20}^{(1)}\}$ for $H_2(S_1)$ and
$\{e_1^{(2)},\ldots,e_{20}^{(2)}\}$ for $H_2(S_2)$ from our
proposition we have volumes of the form
\begin{equation}
\begin{split}
  &\Vol(S_1)\Area(e_a^{(2)})\\
  &\Area(e_a^{(1)})\Vol(S_2).
\end{split}
\end{equation}
The volume of $S_j$ is determined from the K\"ahler form which is
determined by the areas of the $\P^1$'s. Proposition \ref{prop:1} then
implies we have 40 independent functions on 40 variables. If the
superpotential is a suitably generic function then we therefore expect
classical vacua to be isolated in the K\"ahler moduli space. That is,
{\em we fix all the moduli}.

There are two known effects that can spoil the genericity of an
instanton contribution and make it vanish. Firstly, the instanton may
have a moduli space of vanishing Euler characteristic in some
sense. This is not true in our case as rational curves in K3 surface
are always isolated. The second effect can be caused by fluxes
\cite{RS:barren} as we now discuss.

\subsection{Obstructed Instantons}  \label{ss:obs}

Let $D$ be a threefold corresponding to a potential instanton
$S_1\times\P^1$ or $\P^1\times S_2$. Without loss of generality, we
assume the instanton is of the form $C_1\times S_2$ from now on, with
$C_1\cong\P^1$. Let $i:D\hookrightarrow S_1\times S_2$ be the
embedding. The term
\begin{equation}
\int_D b_2\wedge i^*G,
\end{equation}
in the M5-brane worldvolume action induces a tadpole for the
anti-self-dual 2-form $b_2$ if $i^*G\neq0$. We therefore demand that
$i^*G=0$ is a necessary condition for any divisor $D$ to be considered
an instanton.

How strong is the constraint $i^*G=0$? Let us first consider the
supersymmetry-breaking part of the flux
$G_1=\Re(\Omega_1\wedge\overline{\Omega}_2)$. Viewing $G\in
H^4(S_1\times S_2,\Z)$ as a homomorphism from chains in $S_1\times
S_2$ to $\Z$, we may write
\begin{equation}
  i^*G(x) = G(i(x)),
\end{equation}
where $x$ is a 4-chain on $D$. Purely on dimensionality grounds, from
(\ref{eq:Ggen}), it is easy to see that, if $G(i(x))\neq0$, then $x$
must be mapped under $i$ to a 2-chain on $S_1$ and a 2-chain on
$S_2$. We therefore suppose that $i(x)\cong C_1\times C_2$, for some
2-cycle $C_2\subset S_2$. But then $G(i(x))=0$ since $\Omega_1$ is of
type $(2,0)$ and therefore must vanish on any $\P^1$ (as the latter is
dual to a (1,1)-form). This means that {\em none\/} of our instantons
are ruled out by this part of the $G$-flux.

Now let us consider the case where $G_0$ is nonzero and given by
(\ref{eq:G0G1}). These fluxes will fix some of the 20 K\"ahler
moduli. The primitivity condition for $G$ means that $J_j$ will be a
valid K\"ahler form for $S_j$ only if $J_j$ is perpendicular all the
$\omega_j^{(\alpha)}$'s.\footnote{Here we have mentioned only the real
  K\"ahler form. The complex partner of the K\"ahler form is similarly
  obstructed as discussed in \cite{me:mflux}, for example.}
Let is denote this space of K\"ahler forms $V_j^0\subset
H^2(S_j,\R)$. That is,
\begin{equation}
  V_j^0 = \bigcap_\alpha {\omega_j^{(\alpha)}}^\perp,  \label{eq:V0}
\end{equation}
where the perpendicular complement is taken with respect to
$\omega_j^{(\alpha)}$ in the 20 dimensional space
$\Pic(S_j)\otimes\R$. 

Such a nonzero $G_0$ will also rule out certain
instantons. Consider a 4-cycle $x\cong C_1\times C_2$, where both
$C_j$'s are rational curves in $S_j$ and let $\xi_j$ denote the
Poincar\'e dual of $C_j$.  Then
\begin{equation}
G(i(x)) = \sum_\alpha (\omega_1^{\alpha}.\xi_1)(\omega_2^{\alpha}.\xi_2)
\end{equation}
The instanton $C_1\times S_2$ is therefore only valid (i.e., $i^*G=0$)
if $\xi_1$ is orthogonal to all the $\omega_1^{(\alpha)}$'s. That is,
\begin{equation}
  \xi_1 \in  V_1^0.
\end{equation}

Our instantons only contribute nontrivially to the superpotential if
they correspond to $\P^1\times\KIII$, where we assume the $\P^1$ is
holomorphically and smoothly embedded in the K3 surface. That is, the
$\P^1$ is a rational curve. Fortunately these rational curves can be
categorized using properties of the lattice at hand. Any rational
curve in a K3 surface has self-intersection $-2$. This means it is
Poincar\'e dual to an element of the lattice $H^2(S_j,\Z)$ of length
squared $-2$. Conversely, if $\xi$ is an element of length squared $-2$
in $H^2(S_j,\Z)$ then either $\xi$ or $-\xi$ is Poincar\'e dual to a
rational curve. 

This leads to the following:
\begin{theorem}
If $G_0$ is zero we generically fix all moduli. With a nonzero $G_0$,
instanton effects will generically fix all moduli if, and only if, the
spaces $V_1^0$ and $V_2^0$ defined in (\ref{eq:V0}) are spanned by
elements corresponding to rational curves. That is the $V_j^0$'s are
spanned by elements in $V_j^0\cap H^2(S_j,\Z)$ of length squared $-2$.
\label{th:fix2}
\end{theorem}

In simple cases, all the moduli are fixed. For example, suppose $M=1$
in (\ref{eq:G0G1}) and $(\omega_1^{(1)})^2=-2$. Suppose further that
that Picard lattice contains a copy of the (negated) $E_8$ lattice as
a summand and that $\omega_1^{(1)}$ is an element of this
lattice. Then the orthogonal complement of this vector will be the
$E_7$ lattice which is generated by vectors of length squared
$-2$.

It would be interesting to find examples where the moduli are {\em
  not\/} all fixed by instanton effects. This would involve analyzing
  sublattices in $H^2(S,\Z)$ which are not generated by vectors of
  length squared $-2$.

\subsection{The Orientifold Limit}  \label{ss:or2}

By going to the F-theory limit we may obtain the equivalent statement
about instanton effects in the orientifold on
$\KIII\times(T^2/\Z_2)$. Begin with M-theory on $S_1\times S_2$, where
$S_2$ is an elliptic fibration. Let the area of the generic elliptic
fibre be $A$. To take the F-theory limit we set $A\to 0$.

The rescaling involved in this limit means that the volume of the
M5-brane instanton must scale as $A$, as $A\to0$, in order that this
instanton has a nontrivial effect \cite{Wit:super}. It follows that
the instanton must either wrap an elliptic fibre, or a component of a
bad fibre.

The instantons corresponding to $\P^1\times S_2$ indeed wrap the fibre
and so descend to D3-brane instantons wrapped around $\P^1\times(T^2/\Z_2)$
in the F-theory limit.
The instantons corresponding to $S_1\times\P^1$ will be trivial unless
the $\P^1$ corresponds to a component of a bad fibre. In this case,
the D3-brane instanton becomes wrapped on $S_1\times{\mathrm{pt}}$.

The moduli fixing then proceeds in the same way as it did for
M-theory. Unless an inauspicious choice of $G$-flux is used, all the
moduli should be fixed by instanton effects as follows. After flux was
applied, the single remaining modulus in $\cM_V$ corresponded to the
volume of $S_1$. Clearly this is fixed by the instantons wrapping
$S_1\times{\mathrm{pt}}$. The remaining moduli correspond to the areas
of rational curves in $S_1$ and the area of $T^2/\Z_2$. Given that the
volume of $S_1$ has been fixed, we have precisely the right number of
independent constraints from the $\P^1\times(T^2/\Z_2)$ instantons to
fix these latter moduli.

The fact that the single vector multiplet corresponding to the volume
of $S_1$ is fixed was also observed in \cite{BM:K3m}, where a more
quantitative analysis was performed using duality.


\section{Discussion} \label{s:conc}

If one considers M-theory on $\KIII\times\KIII$
with no M2 branes and a flux chosen to break supersymmetry down to
$N=2$ in three dimensions, then the complex structures of the two K3
surfaces are fixed. To be precise, the two K3 surfaces are both
attractive K3 surfaces. There remain 20 complex moduli associated to
each K3 surface which vary the K\"ahler form.

If we leave the 40 moduli unfixed by fluxes, then we have argued that
generically one would expect instanton effects to fix all 40. If flux
is used to fix further moduli then we showed that there is a
possibility that some moduli can remain unfixed by instanton effects.

The obvious next step should be to compute these instanton effects
more explicitly and determine the values of the moduli. This might be
a difficult exercise for the following reasons.

Before the flux was turned on we have an $N=4$ supersymmetric theory
in three dimensions. Corrections from M5-brane instantons will effect
the metric on the moduli space.  The moduli space of this theory is a
product of quaternionic K\"ahler moduli spaces. It is a well-known
difficult problem in string theory to determine the form of such
quaternionic K\"ahler moduli spaces when there are nontrivial
instanton corrections. The problem of studying M5-brane instantons
corrections to the moduli space is exactly equivalent to studying
worldsheet instanton corrections to the heterotic string on a K3
surface. Preliminary analysis in the latter was done in
\cite{W:K3inst}, for example, but few concrete results have been
attained.

It should be emphasized that, even though there has been much
interesting progress on quaternionic K\"ahler manifolds (such as
\cite{FS:quat,D'Auria:2004tr}), these results tend to rely on the
assumption that there is an isometry in the moduli space related to
translations in the RR directions. This views the hypermultiplet
moduli space as a fibration over some special K\"ahler base with a
toroidal fibre given by the RR moduli. It is known (see
\cite{AD:tang}, for example) that when non-perturbative corrections
are taken into account, this fibration must have ``bad fibres''. These
bad fibres will break these isometries in much the same way as an
elliptic K3 surface has no isometries related to translation in the
fibre direction. An interesting proposal for analyzing instanton
effects on the hypermultiplet moduli space was given in \cite{BM:K3m}
but it appears to rely on the existence of these isometries.

Now when we turn the flux on, the M5-brane instantons contribute to a
superpotential, rather than the moduli space metric. This does not
mean that the metric remains uncorrected however. Now, with the
decreased supersymmetry, quantum corrections to the metric are less
constrained and even more difficult to determine than if they arose
purely from instantons.  We see, therefore, that computing the
superpotential directly from instanton computations may be very
difficult.

Even without this detailed
knowledge, however, we have shown that one should expect a number of
flux compactifications associated to M-theory on $\KIII\times\KIII$
(or its equivalent orientifold $\KIII\times(T^2/\Z_2)$) where all the
moduli are fixed by the combined action of the flux and the instanton effects. 

In the context of the F-theory limit, which is equivalent to an
orientifold of the type IIB string on $\KIII\times (T^2/\Z_2)$ the
result of this paper shows that the goal of fixing all moduli in this
model is now accomplished.\footnote{In \cite{BKKST:D3} the counting of
fermionic zero modes on D3 brane is performed which leads to an
analogous result.} The first part, namely fixing the moduli by fluxes,
was achieved in \cite{TT:K3T2,Andrianopoli:2003jf} and a nice summary
of this work was presented in \cite{D'Auria:2004td}. In absence of
D3-branes, the 18+1 complex moduli ``unfixable'' by fluxes span the
scalar manifold
\begin{equation}
  \cM_{_{\textrm{unfixed}}}^{^{\textrm{min}}} 
= \frac{\GO(2,18)}{\GO(2)\times\GO(18)} \times
  \frac{\Sl(2,\R)}{\GU(1)}. \label{eq:Fer}
\end{equation}
Here the first factor includes 18 complex fields, the remnant of the
$N=2$ hypermultiplets. It includes the area of $(T^2/\Z_2)$ and other
hypermultiplets. The second factor is the remnant of the $N=2$ vector
multiplet and it describes the volume of the K3 surface $S_2$ and its axionic
partner. This case in our setting requires that both $G_{1}$ and
$G_{0}$ are non-vanishing. We have to be careful therefore and comply
with the conditions of the theorem \ref{th:fix2}, where it is explained that
only certain choice of fluxes $G_{0}$ will allow us enough freedom
(18+1 choice of proper 4-cycles) to fix by the instantons all
remaining 18+1 complex moduli in (\ref{eq:Fer}).
  
An even simpler case, from the perspective of instantons, is when we
introduce only $G_{1}$ flux (breaking $N=2$ into $N=1$ supersymmetry) will
leaves us with the 20+1 complex moduli unfixed by fluxes.  They span
the scalar manifold
  \begin{equation}
  \cM_{_{\textrm{unfixed}}} = \frac{\GO(2,20)}{\GO(2)\times\GO(18)} \times
  \frac{\Sl(2,\R)}{\GU(1)}. \label{eq:Fer1}
\end{equation}
In such case we simply have 20+1 choices for the D3 instantons
wrapping the 4-cycles in $\KIII\times (T^2/\Z_2)$ and all unfixed by
fluxes moduli are fixed by instantons.
  
The whole story of fixing all moduli in the M-theory version of this
model, compactified on $\KIII\times \KIII$ is incredibly simple and
elegant. In the compactified three-dimensional model there are no
vectors. Therefore without fluxes, we have two 80-dimensional
quaternionic K\"ahler spaces, one for each $\KIII$. With non-vanishing
$G_{1}$ flux, each $\KIII$ becomes an attractive $\KIII$, one-half of
all the moduli are fixed, but 40 in each $\KIII$ still remain moduli and
need to be fixed by instantons. There are 20 proper 4-cycles in each
$\KIII$ and they provide instanton corrections from M5-branes wrapped
on these cycles: the moduli space is no more.


\section*{Acknowledgments}

We wish to thank N.~Bliznashki, S.~Ferrara, L.~Fidkowski, C.~Haase,
S.~Kachru, A.-K.~ Kash\-ani-Poor, D.~Kraines, D.~Morrison, G.~Moore,
S.~Sethi, A.~Tomasiello and S.~Trivedi for useful
conversations. P.S.A.~is supported in part by NSF grant DMS-0301476,
Stanford University, SLAC and the Packard Foundation.  R.K.~is
supported by NSF grant 0244728.


\end{document}